\documentclass[aip, prb, reprint, amsmath, amssymb, nofootinbib]{revtex4-2}
\usepackage[utf8]{inputenc}
\usepackage{setspace}
\usepackage{booktabs}
\usepackage[dvips]{graphicx}
\usepackage[dvipsnames]{xcolor}
\definecolor{CiteBlue}{RGB}{45,52,151}
\usepackage[
    colorlinks=true,
    linkcolor=CiteBlue,
    urlcolor=CiteBlue,
    citecolor=CiteBlue
]{hyperref}
\usepackage[capitalise]{cleveref}
\usepackage{tikz}
\usepackage{siunitx}
\DeclareSIUnit{\year}{yr}
\DeclareSIUnit{\erg}{erg}

\newcommand{\refcite}[1]{Ref.~\onlinecite{#1}}
\newcommand{\refscite}[1]{Refs.~\onlinecite{#1}}

\usepackage{bm}
\newcommand{\bb}[1]{\bm{\mathrm{#1}}}
\newcommand{\du}{\mathrm{d}}
\newcommand{\dd}{\,\du}

\begin{document}
\title{Higher multipoles of the cow}
\author{Benjamin V. Lehmann}
\affiliation{Center for Theoretical Physics---a Leinweber Institute, Massachusetts Institute of Technology, Cambridge, Massachusetts 02139, USA}
\date{1 April 2026}
\begin{abstract}
The spherical cow approximation is widely used in the literature, but is rarely justified. Here, I propose several schemes for extending the spherical cow approximation to a full multipole expansion, in which the spherical cow is simply the first term. This allows for the computation of bovine potentials and interactions beyond spherical symmetry, and also provides a scheme for defining the geometry of the cow itself at higher multipole moments. This is especially important for the treatment of physical processes that are suppressed by spherical symmetry, such as the spindown of a rotating cow due to the emission of gravitational waves. I demonstrate the computation of multipole coefficients for a benchmark cow, and illustrate the applicability of the multipolar cow to several important problems.
\end{abstract}
\maketitle

\clearpage

\section{Introduction}
\label{sec:intro}

Various forms of the spherical cow approximation (SCA) underlie crucial results across an enormous range of subfields. A Google Scholar search indicates that spherical cows can be found in over 1000 published papers~\cite{scholar}. The SCA is so pervasive that it has even become one of the few approximation methods that we regularly teach to first-year undergraduates. Given the importance of the SCA in physics, it is shocking and alarming that no quantitative assessment of its validity has ever appeared in the literature.

Moreover, the SCA is impoverished as an approximation technique because it is not systematically improvable in its usual form. In most approximation schemes, while one might work at leading order, it is possible in principle to consider contributions from next-to-leading order, or higher orders, whether to improve the result or to assess convergence of the approximation. But a spherical cow is simply a ball, with no room for additional parameters to improve the matching to a realistic cow. This should leave us all feeling quite sheepish.

However, this is not intrinsic to the structure of the SCA itself. Given an appropriate scheme for higher-order corrections, the spherical cow could be treated as the leading-order term of a well-formulated expansion. Indeed, there is a standard approximation technique that lends itself well to this purpose: the spherical multipole expansion. The first term of the multipole expansion---the monopole---is spherically symmetric, and higher-order terms encode deviations from spherical symmetry on progressively smaller scales. If the spherical cow can be reinterpreted as the monopole of a multipole expansion, then the SCA becomes the first term in a systematically improvable approximation, allowing for a quantitative test of its reliability in various settings.

In this work, I introduce several different schemes for implementing this expansion and extending the SCA to include higher multipoles of the cow. I directly compare the contribution of higher multipoles to the monopole (i.e., the spherical cow), performing the first true test of the validity of the SCA under various circumstances. I use this framework to identify cases where the SCA is clearly insufficient, and in which dipole or higher multipole contributions must be included to obtain physically realistic results even at the order-of-magnitude level. As I explain, even some of the most classic bovine problems, such as the cow tipping problem, receive dominant contributions from higher multipoles.

Since the SCA itself takes several different forms, the extension to higher multipoles also varies by use case. Here I focus on two cases. In \cref{sec:multipole-potential}, I study the inclusion of higher multipoles in potentials sourced by the cow, and discuss some significant consequences for processes that are suppressed by spherical symmetry, including gravitational radiation. In \cref{sec:bovine-geometry}, I define a multipole expansion for the geometry of the cow itself, which allows for more general extensions of the SCA\@. This latter case is especially important for problems in bovine rigid body mechanics (e.g., cow tipping), which are treated in \cref{sec:rigid-body}. I discuss the implications of these results and conclude in \cref{sec:discussion}.

Throughout this work, for numerical computations, I use a benchmark cow from the \texttt{libigl} tutorial data~\cite{libigl,libigl-cow}. I denote the interior of the cow by $\mathcal C$, and the 2-dimensional surface of the cow by $\partial\mathcal C$. I follow IUPAC conventions for multiplier prefixes~\cite{IUPAC:2014}, so, for example, terms with $\ell = 256$ are said to be components of the hexapentacontadictapole moment.

\section{Higher multipoles of bovine potentials}
\label{sec:multipole-potential}
Let us first consider the simplest definition of the spherical cow, as it appears in the context of gravitational problems. Of course, the same methods apply to the study of the electrostatic potential in the case of a charged cow. However, charging cows are rarely observed in nature relative to charging bulls~\cite{lott:1981,douglass:1999}. (Still, caution is advised in experimental settings, per \refcite{murphy:2010}.)

When computing the gravitational potential $\phi(\bb x)$ at a point $\bb x$ outside the cow, the solution of the Laplace equation can be expanded in spherical harmonics $Y_\ell^m(\bb{\hat x})$ via the irregular solid harmonics, which yields the multipole expansion
\begin{equation}
    \label{eq:spherical-multipole-potential}
    \phi(\bb x) = -G\sum_{\ell = 0}^{\infty}
        \frac{
            \left[\frac{4\pi}{2\ell+1}\right]^{1/2}
        }{\|\bb x\|^{\ell+1}}
        \sum_{m=-\ell}^\ell (-1)^m Y_\ell^{-m}(\bb{\hat x})
        Q_\ell^m
        ,
\end{equation}
where the spherical multipole moment $Q_\ell^m$ is defined by
\begin{equation}
    Q_\ell^m\equiv \int_{\mathcal C}\dd^3\bb x\,
    \rho(\bb x)\,\|\bb x\|^\ell Y_\ell^m(\bb{\hat x})
    ,
\end{equation}
given a mass density $\rho(\bb x)$. (See Ref.~\cite{Jackson:1998nia} for a thorough derivation.) I normalize the spherical harmonics as
\begin{equation}
    Y_\ell^m(\bb{\hat x}) = \sqrt{
        \frac{(2\ell + 1)!}{4\pi}\frac{(\ell-m)!}{(\ell+m)!}
    }\,P_\ell^m(\cos\theta)\,e^{im\theta}
    ,
\end{equation}
where $P_\ell^m$ denotes the associated Legendre polynomials, and $\bb{\hat x}$ is the unit vector in the direction of $\bb x$. I will take $\rho$ to be constant over the cow, and I will work in ``cow coordinates,'' where the $x$ axis is aligned with the forward direction of the cow, the $y$ axis is orthogonal to the ground, and the positive $z$ axis points to the cow's right. This matches the coordinate axes of the benchmark cow. I also retain the scaling of the benchmark cow, so the length units correspond to a bounding box for the cow with dimensions $(1.044, 0.6397, 0.3403)$ in $x$, $y$, and $z$, respectively. I call these units ``benchmark units,'' with conversions to SI units provided later in this section. The first few multipole moments for the benchmark cow are given in these units in \cref{tab:mass-moments}. The monopole corresponds to the spherical cow, so the higher multipole coefficients indicate the size of corrections to the SCA\@.

\begin{table*}\centering
    \caption{Leading multipoles of the cow $\mathcal C$ treated as a mass distribution, computed about the center of mass in benchmark units. In barycentric coordinates, the dipole moment vanishes, so these components are omitted. Additionally, since $Q_\ell^{-m} = (-1)^m Q_\ell^{m*}$, only the positive-$m$ coefficients are shown.}
    \setlength{\tabcolsep}{10pt}
    \renewcommand{\arraystretch}{1.15}
    \setlength{\heavyrulewidth}{0.8pt}
    \setlength{\lightrulewidth}{0.5pt}
    \setlength{\cmidrulewidth}{0.5pt}
    \setlength{\aboverulesep}{0.4ex}
    \setlength{\belowrulesep}{0.4ex}
    {\footnotesize
    \begin{tabular*}{\textwidth}{@{\extracolsep{\fill}}ccc@{\hspace{2.5em}}ccc@{}}
    \addlinespace[1ex]
    \toprule
    \addlinespace[-0.85ex]
    \toprule
    \addlinespace[1.5ex]
        $\ell$ & $m$ & $Q_\ell^m$ & $\ell$ & $m$ & $Q_\ell^m$ \\
    \addlinespace[0.2ex]
    \midrule
    \addlinespace[1ex]
        0 & 0 & 0.0152
            & 3 & 3 & $-0.0001 + 0.0001i$ \\
        2 & 0 & $-0.0008$
            & 4 & 0 & 0.0001 \\
        2 & 1 & $\num{-1.81e-06} + \num{1.01e-07}i$
            & 4 & 1 & $\num{1.58e-07} + \num{3.94e-07}i$ \\
        2 & 2 & $0.0008 - 0.0002i$
            & 4 & 2 & $-0.0001 + \num{4.00e-05}i$ \\
        3 & 0 & \num{-1.63e-06}
            & 4 & 3 & $\num{-4.01e-07} + \num{1.51e-07}i$ \\
        3 & 1 & $0.0001 - \num{3.42e-05}i$
            & 4 & 4 & $\num{3.16e-05} - 0.0001i$ \\
        3 & 2 & $\num{-2.83e-07} - \num{6.81e-07}i$
            & 5 & 0 & \num{2.97e-07} \\
    \bottomrule
    \addlinespace[0ex]
    \bottomrule
    \end{tabular*}
    }
    \label{tab:mass-moments}
\end{table*}

We can now use these coefficients to study gravitational phenomena in the bovine potential. There are several potential applications. For instance, the quadrupole encodes the equivalent of Earth's equatorial bulge for a heavy cow. Additionally, a cow with a nonvanishing quadrupole moment will experience a torque in a nonuniform gravitational field, so a cow falling from a high altitude in Earth's gravitational field will favor a particular orientation. Thus, the quadrupole moment is crucial to answering the question of whether a dropped cow tends to land on its feet. For the same reason, the quadrupole plays a key role in the tidal locking of cows in orbit.\footnote{A similar computation would apply for pigs in orbit, but these are only expected to be observed when pigs fly.} Neither of these cases is amenable to experimental study, and in each, the SCA clearly fails.

However, for the moment, let us focus on another application of the quadrupole moment: computation of the rate of emission of gravitational radiation. A freely rotating cow in vacuum will slow down as it loses energy and angular momentum to gravitational waves. We can compute the rate of spindown using the quadrupole formula for gravitational radiation. When written in terms of the components of the spherical quadrupole moment, this reads~\cite{Maggiore:2007ulw}
\begin{equation}
    \dot E_{\mathrm{quad}} =
    \frac{3G}{8\pi c^5}\sum_{m=-2}^2
    \left\langle\left|\dddot{Q}_2^m\right|^2\right\rangle
    .
\end{equation}
In principle, one could take the spherical multipole coefficients from \cref{tab:mass-moments}, transform them under rotations, and use this transformation behavior to evaluate $\dddot{Q}_2^m$ for a rotation about a given axis. It is simpler, however, to work with Cartesian multipole coefficients for this purpose, since the Cartesian quadrupole coefficients take the form of a symmetric tensor with components $Q^{\mathrm{C}}_{ij} = \int\du^3\bb x\,\rho(\bb x)\,(3x_ix_j - r^2\delta_{ij})$. The Cartesian quadrupole tensor of the benchmark cow in cow coordinates and benchmark units with $\rho(\bb x) = 1$ is
\begin{equation}
    \label{eq:cartesian-quadrupole}
    Q^{\mathrm{C}} =
    \begin{pmatrix}
        4.23004 &  \phantom{-}0.83531 &  \phantom{-}0.00704 \\
        0.83531 & -1.64753 &  \phantom{-}0.00039 \\
        0.00704 &  \phantom{-}0.00039 & -2.58252
    \end{pmatrix}
    \times\num{e-3}
    ,
\end{equation}
and the radiated power is given in terms of the $Q_{ij}^{\mathrm{C}}$ by
\begin{equation}
    \dot E_{\mathrm{quad}} = \frac{G}{45c^5}
        \left\langle\dddot{Q}^{\mathrm{C}}_{ij}\dddot{Q}^{\mathrm{C}\,ij}
        \right\rangle
    .
\end{equation}
Now, suppose the cow is rotating about an axis $\bb{\hat a}$ with angular frequency $\omega$. If $R_{ij}(\theta)$ denotes the rotation matrix about $\bb{\hat a}$ by an angle $\theta$, then at any time $t$, we can write the quadrupole tensor as 
\begin{equation}
    Q_{ij}^{\mathrm{C}}(t) = R^{ik}(\omega t)\,R^{j\ell}(\omega t)\,
        Q_{k\ell}^{\mathrm{C}}(0)
    ,
\end{equation}
so the third derivative in time is given by
\begin{equation}
    \label{eq:quadrupole-derivative}
    \dddot{Q}_{ij}^{\mathrm{C}}(t) = 
        Q_{k\ell}^{\mathrm{C}}(0)\,
        \partial_t^3\left[R^{ik}(\omega t)\,R^{j\ell}(\omega t)\right]
    .
\end{equation}
For a simple example, consider a constant rotation about the $y$ axis, as might be experienced by a cow wearing misaligned rollerskates. Here the rotation matrix is
\begin{equation}
    R(\theta) =
    \begin{pmatrix}
        \cos\theta & 0 & \sin\theta \\
        0 & 1 & 0 \\
        -\sin\theta & 0 & \cos\theta
    \end{pmatrix}
    .
\end{equation}
Inserting this into \cref{eq:quadrupole-derivative} and averaging the result over a full period gives $\langle\dddot{Q}^{\mathrm{C}}_{ij}\dddot{Q}^{\mathrm{C}\,ij}\rangle \approx 0.00149 \omega^6$.

We can now restore physical dimensions to this result, which has units of $M^2L^4\omega^6$, where $M$ and $L$ are the units of mass and length for the benchmark cow. The average length of a cow is about \qty{2.5}{\meter}, and the length in benchmark units is 1.044, so let us say that $L=\qty{2.39}{\meter}$. To determine the mass unit, recall that we set $\rho(\bb x) = 1$, and the actual density of a cow is about \qty{1}{\gram/\centi\meter^3}. Setting $M/L^3 = \qty{1}{\gram/\centi\meter^3}$ gives $M = \qty{13652}{\kilo\gram}$. Note that this is \textit{not} the mass of the cow itself, but simply the mass unit implied by our choice to set particular physical values of length and density to 1. (The volume of the cow in benchmark units is only 0.054, corresponding a physical mass of about \qty{740}{\kilo\gram}---quite typical for a cow.) Replacing these quantities, in physical units, we obtain $\langle\dddot{Q}^{\mathrm{C}}_{ij}\dddot{Q}^{\mathrm{C}\,ij}\rangle \approx \qty{9e12}{\kilo\gram^2\meter^4}\times\omega^6$, or
\begin{equation}
    \dot E_{\mathrm{quad}} \approx \qty{5.5e-41}{\erg/\second} \times
        \left(\frac{\omega}{\qty{1}{\hertz}}\right)^6
        .
\end{equation}

To understand the relevance of this energy loss rate, we must first compute the energy of rotation. Recall that the rotational energy at an angular frequency $\omega$ is given by $E = \frac12I\omega^2$, where $I$ is the moment of inertia about the axis of rotation. For a rotation about an arbitrary axis, this generalizes to a vectorial expression: the moment of inertia is replaced by the inertia tensor, with components $I_{ij} = \int\du^3\bb x\,\rho(\bb x)\,\left(r^2\delta_{ij} - x_ix_j\right)$. (Note that this is similar but not identical to the Cartesian quadrupole tensor of \cref{eq:cartesian-quadrupole}: the Cartesian multipole tensors originate from a Taylor expansion of the potential, which introduces different constant factors multiplying the Cartesian coordinates at each order in the expansion.) The inertia tensor in benchmark units is
\begin{equation}
    I =
    \begin{pmatrix}
         \phantom{-}7.95079 & -2.78437 & -0.02348\\
        -2.78437 &  \phantom{-}27.5426 & -0.00130\\
        -0.02348 & -0.00130 &  \phantom{-}30.6593
    \end{pmatrix}
    \times\num{e-4}
    ,
\end{equation}
so we can readily determine the kinetic energy associated with rotation about the $y$ axis. (We can also see that at fixed $\omega$, the most energetic spinning would involve rotation about the $z$ axis, i.e., falling head over heels.) We can thus obtain the timescale on which the rotation slows (neglecting the details of radiation of angular momentum):
\begin{equation}
    \frac{E}{\dot E_{\mathrm{quad}}} \approx \qty{1.9e49}{\second}
    \left(\frac{\omega}{\qty{1}{\hertz}}\right)^{-4}
    .
\end{equation}
This implies that a cow that began spinning with an angular frequency of \qty{1}{\hertz} at \qty{12}{\mega\year} after the Big Bang\footnote{This corresponds to a cosmic temperature resulting in a medium-well done steak. For a rare steak, begin at \qty{13}{\mega\year}.} would, at the present day, spin at a paltry rate of
\begin{widetext}
\begin{equation}
    \omega_{\mathrm{now}} = \qty{0.999999999999999999999999999999988568}{\hertz}.
\end{equation}
\end{widetext}
Since there is no gravitational radiation due to the monopole or dipole moments, this effect is completely lost when the mass distribution is regarded as purely monopolar. Thus, this is another case in which the SCA is wholly inadequate---under the SCA, one would incorrectly conclude that the cow never slows its spinning!

Again, I stress that while the spherical cow naturally corresponds to the spherically-symmetric monopole term, the monopole term actually represents the potential of a point mass, not a ball. The multipole expansion of the potential does not readily encode the full geometry of the source. In particular, the expansion is only well defined outside the source distribution---even including arbitrarily many terms, the potential of \cref{eq:spherical-multipole-potential} is sourceless everywhere away from the origin. We will study corrections to the spherical cow's geometry in the next section.

\begin{figure}
    \includegraphics[width=0.45\textwidth]{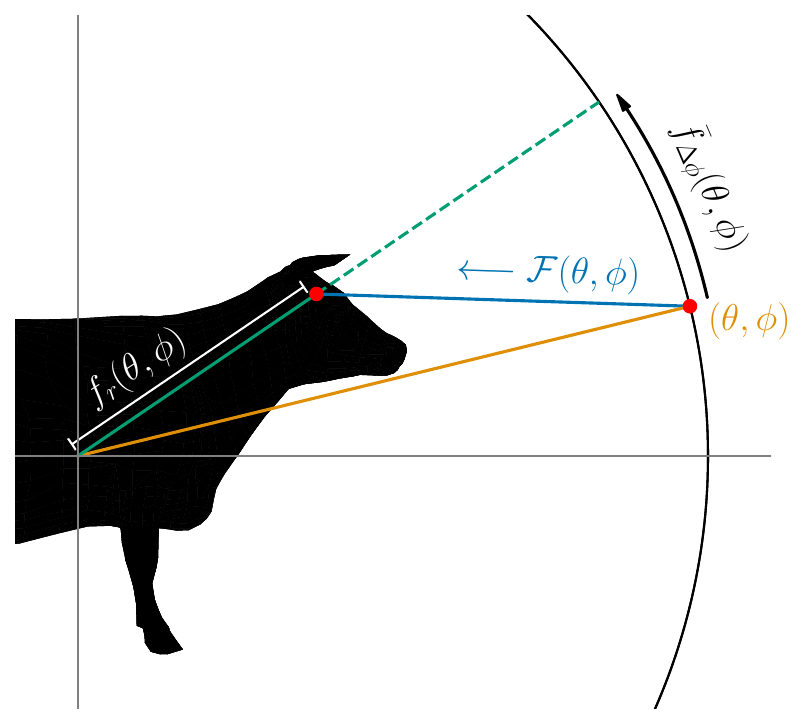}
    \caption{Illustration of the choice of $\bb f$ used in this work (projected). The point $(\theta, \phi)\in S^2$ is mapped by the homeomorphism $\mathcal F$ to a point on the surface of the cow. This point, $\mathcal F(\theta, \phi)\in\partial\mathcal C$, has a radial coordinate $f_r(\theta, \phi)$, and also has different angular coordinates from the original point. The differences in the $\theta$ and $\phi$ coordinates are given by $\bar{f}_{\Delta\theta}$ and $\bar{f}_{\Delta\phi}$, respectively. The means of these angular differences are then subtracted by rotation of the coordinate system.}
    \label{fig:delta-f-bar}
\end{figure}

\begin{figure*}\centering
    \begin{tikzpicture}
        \node at (0, 0)
            {\includegraphics[height=5cm]{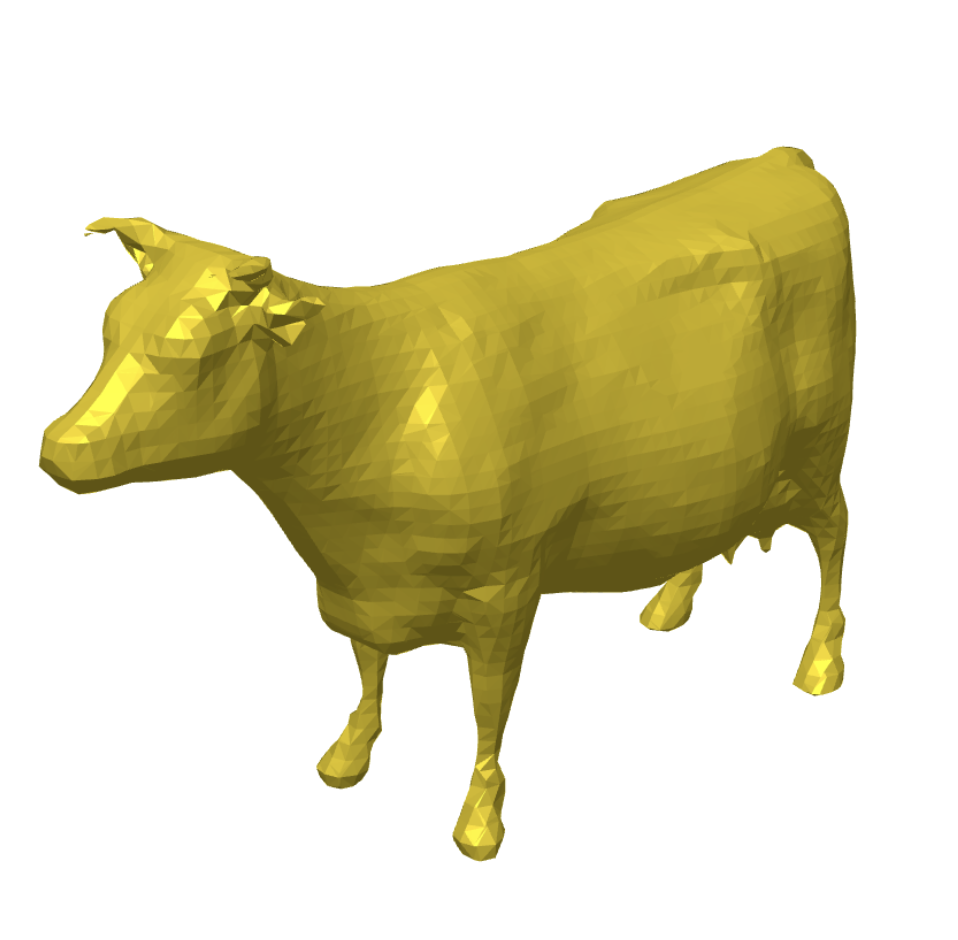}};
        \node at (6, 0)
            {\includegraphics[height=5cm]{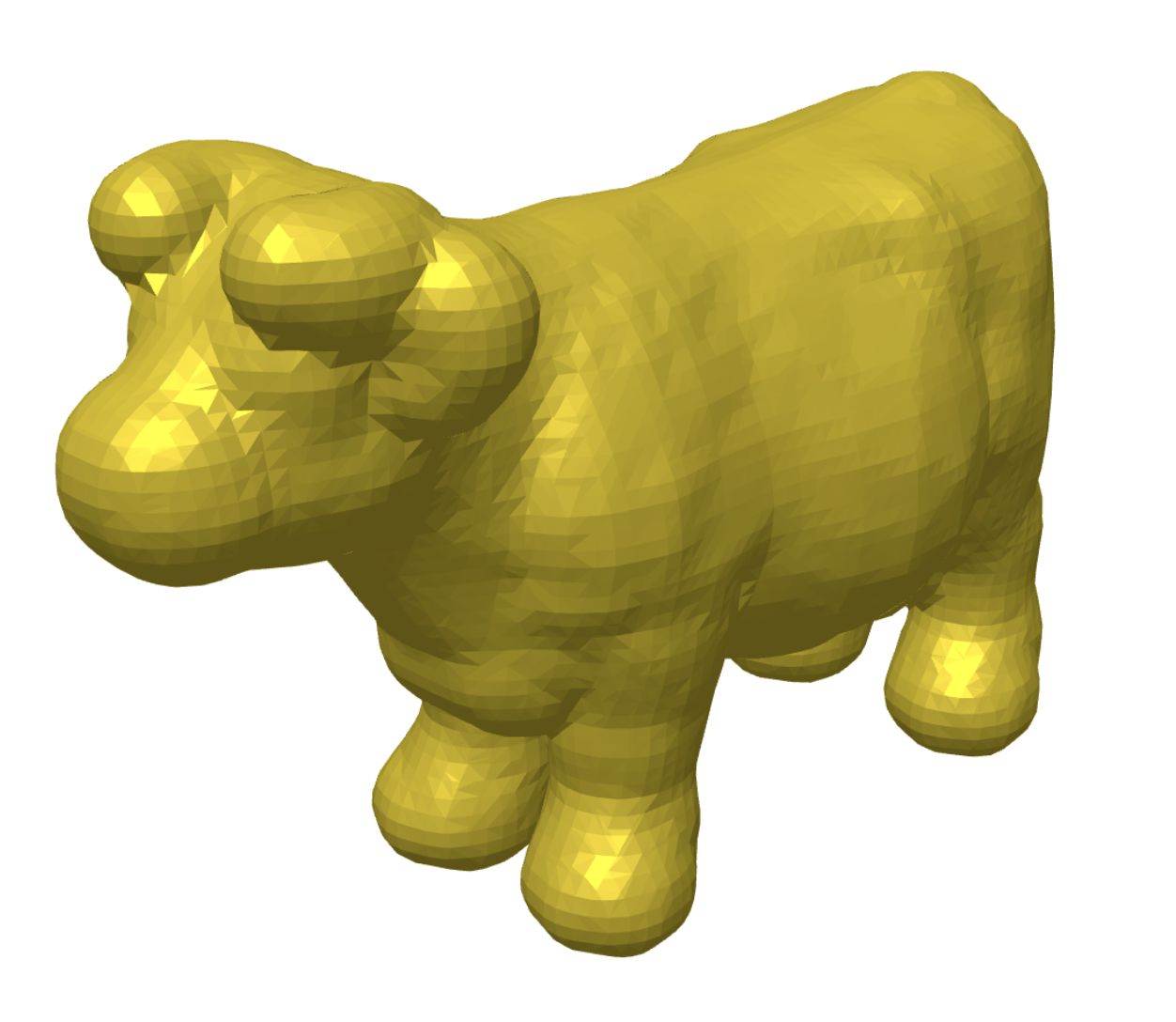}};
        \node at (12, 0)
            {\includegraphics[height=5cm]{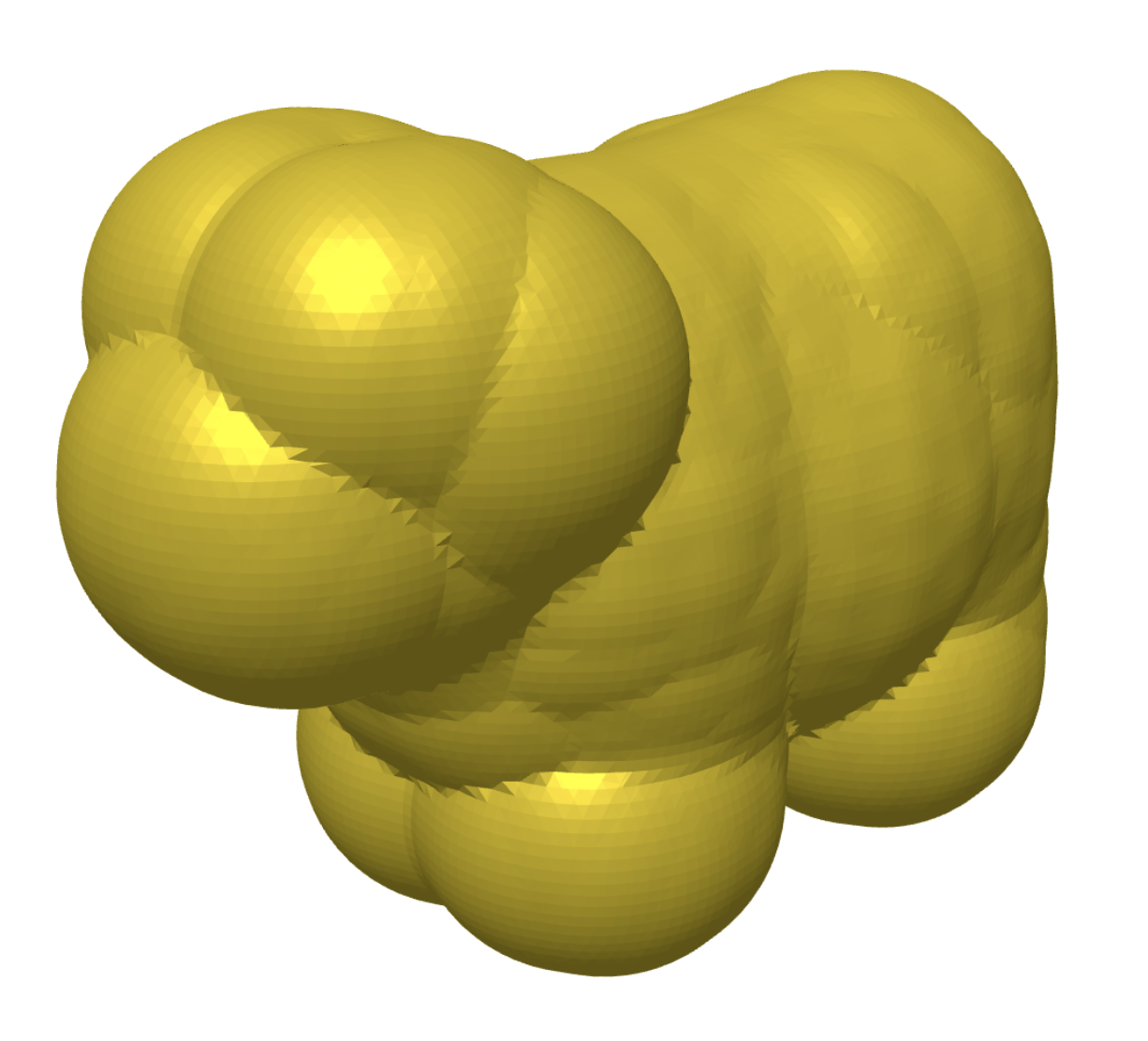}};
        \draw[->, thick] (0, 1.5) to [bend left]
            node[above] {$\bb\varphi(\cdot\,, \Delta t)$} ++(3.5, 0.5);
        \draw[->, thick] (6, 2)  to [bend left]
            node[above] {$\bb\varphi(\cdot\,, \Delta t)$} ++(4, 0);
        \draw[->, thick] (5, -2) to [bend left]
            node[below] {$\bb\varphi(\cdot\,, -\Delta t)$} ++(-4, 0.5);
        \draw[->, thick] (11, -2) to [bend left]
            node[below] {$\bb\varphi(\cdot\,, -\Delta t)$} ++(-4, 0.5);
        \coordinate (o) at (-1.5, -2);
        \draw[->, shorten >=2mm] (o) -- ++(0, 0.7)
            node[font=\footnotesize] {$y$};
        \draw[->, shorten >=2mm] (o) -- ++(-0.6, -0.25)
            node[font=\footnotesize] {$x$};
        \draw[->, shorten >=2mm] (o) -- ++(-0.4, 0.25)
            node[font=\footnotesize] {$z$};
        \fill[black] (o) circle (2pt);
    \end{tikzpicture}
    \caption{Increasing level sets of $d_{\mathcal C}$. The function $\bb\varphi(\bb x, t)$ maps between these level sets from left to right with increasing $t$. This has the effect of ``inflating'' the cow to a star-shaped domain, from which the boundary can be projected onto the sphere smoothly and bijectively.}
    \label{fig:puffy-cow}
\end{figure*}

\begin{figure}
    \includegraphics[width=0.45\textwidth]{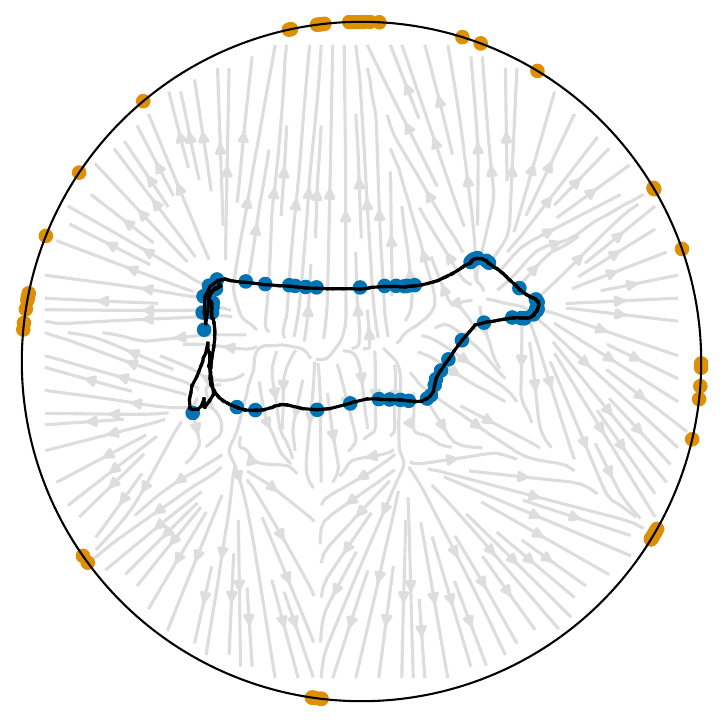}
    \caption{Distance gradient flow as a mapping to the unit sphere ($\mathcal F^{-1}$), shown here for points within 0.001 benchmark units of the $yz$ plane. The original points are shown in the interior (blue). Points on the circle (orange) are obtained by integrating along the gradient shown by the gray streamlines. Note the sharply different behavior of the streamlines above and below the cow: below, the presence of the legs (out of the plane shown) pushes the flow away from the front and back of the cow.}
    \label{fig:distance-gradient-flow}
\end{figure}

\section{Higher multipoles of bovine surfaces}
\label{sec:bovine-geometry}
In the previous section, we used the multipole expansion of the mass distribution and gravitational potential to study bovine phenomenology induced by the nonspherical geometry. This is a paradigmatic example of how a particular problem might be solved beyond the spherical cow approximation, with contributions ordered by their asphericity. However, it gives us little intuition for the shape of the cow itself at successive terms in the series. The monopole term clearly corresponds to the spherical cow due to its spherical symmetry, but what is the shape of the cow that corresponds to inclusion of the dipole or quadrupole? In this section, we answer this question by defining a multipole expansion of the cow's \emph{geometry} based on the multipole expansion of functions on the 2-sphere.

We begin by defining the boundary surface of the cow, $\partial\mathcal C$, in a form that is amenable to a direct multipole expansion. To that end, our goal is to represent $\partial\mathcal C$ in terms of a function on $S^2$ that can be expanded in spherical harmonics. Clearly, a typical cow is topologically equivalent to a sphere,\footnote{Here I ignore the alimentary canal, since it is well known that including this results in pure bovine waste.} so there must exist many smooth deformations (homeomorphisms) $\mathcal F\colon S^2\to\partial\mathcal C$ mapping the sphere onto the surface of the cow. Such a map $\mathcal F$ can be written in terms of a set of maps $f_i\colon S^2\to\mathbb R$ (e.g., the vector components of points in the image), and the $f_i$ admit a representation in spherical harmonics as 
\begin{equation}
    \bb f(\Omega) = \sum_{\ell, m} \bb f_\ell^m(\Omega),
    \qquad
    \bb f_\ell^m(\Omega)\equiv
        \langle \bb f|Y_\ell^m\rangle_{S^2} Y_\ell^m(\Omega)
    ,
\end{equation}
where $\Omega$ denotes the angular coordinates on the sphere, and $\langle\cdot|\cdot\rangle_{S^2}$ denotes the $L^2$ inner product on $S^2$,
\begin{equation}
    \langle u | v\rangle = \int_{S^2}\du\Omega\,uv^*
    .
\end{equation}
(Note that some conventions exchange the order of the functions in the brackets above.) Here, we are simply using the fact that the spherical harmonics provide a complete orthonormal system for real-valued functions on the sphere. The monopole term, $\bb f_0^0(\Omega) = \langle \bb f |Y_0^0\rangle Y_0^0$, is spherically symmetric in the $\bb f$ coordinate system, and can thus be used to define the spherical cow. Higher-order terms give corrections to the cow boundary $\partial\mathcal C$.

While the decomposition of $\mathcal F$ into real-valued component maps $\bb f$ is in principle arbitrary, there is a simple and natural choice. In order to respect the spherical symmetry of the monopole term, let us define $\bb f(\theta, \phi) = (f_r, f_{\Delta\theta}, f_{\Delta\phi})$, where $f_r(\Omega)$ gives the radial component of $\mathcal F(\Omega)$, and $f_{\Delta\theta}$ and $f_{\Delta\phi}$ give the difference in the polar and azimuthal angles produced by application of $\mathcal F$. Specifically, let us define
\begin{equation}
    \bar{f}_{\Delta\theta}(\theta, \phi) \equiv
        [\mathcal F(\theta, \phi)]_\theta - \theta,
\end{equation}
where $[\bb x]_\theta$ denotes the $\theta$ coordinate of the point $\bb x$. In general, there is no guarantee that the mean of $\bar{f}_{\Delta\theta}(\theta, \phi)$ vanishes, but the mean can always be subtracted. Thus, let us choose $f_{\Delta\theta}(\theta,\phi) \equiv \bar{f}_{\Delta\theta}(\theta,\phi) - (4\pi)^{-1/2}\bar{f}_{\Delta\theta}^{(0)}$, where $\bar{f}_{\Delta\theta}^{(0)}$ denotes the monopole coefficient of $\bar{f}_{\Delta\theta}$. This definition of $f_{\Delta\theta}$ effects a rotation of the coordinates such that the monopole of $f_{\Delta\theta}$ vanishes. An analogous definition can be made for $f_{\Delta\phi}$. \Cref{fig:delta-f-bar} illustrates the definition of $\bar{f}_{\Delta\phi}$, since the azimuthal angle is easy to visualize in the $xy$ plane, and $\bar{f}_{\Delta\theta}$ is obtained by performing the same procedure with the polar angle.

This choice of $\bb f$ guarantees that the image of the monopole term, with constant $(r, \Delta\theta, \Delta\phi)$, is truly a sphere, and not merely a point, or a surface homeomorphic to $S^2$, and involves no spurious rotation of the coordinate system. In particular, this means that a perfectly spherical cow is always absorbed entirely by the monopole term of the radial coordinate, and has vanishing higher multipole contributions, as desired.

All that remains is to implement this prescription is to identify the map $\mathcal F$. This is an example of a ``surface matching'' problem, which has been well studied in computational geometry. In the remainder of this section, I will describe two surface matching algorithms that give rise to multipole expansions of the cow surface.

\subsection{Distance gradient flow method}
First, let us solve the surface matching problem in a geometrically-intuitive way. The method follows from the realization that if the cow were simply rounder, every point on the boundary would have a unique angular coordinate, meaning that it would clearly be trivial to project onto the sphere. For such a round cow, $\partial\mathcal C$ could by defined by its radial coordinate via a single function $r_{\partial\mathcal C}\colon S^2\to\mathbb R$, which would then admit a multipole expansion. While this is not possible for a general cow, any cow can be first fattened up into a rounder shape and then projected to the sphere, which defines the map $\mathcal F^{-1}$.

More precisely, suppose there exists an automorphism $\bb\varphi\colon\mathbb R^3\to\mathbb R^3$ such that $\bb\varphi(\mathcal C)$, the image of the cow, is a star-shaped domain. This means that it is possible to choose an origin point, $\bb 0$, such that for every point $\bb x\in\bb\varphi(\mathcal C)$, the line segment from $\bb 0$ to $\bb x$ lies entirely within $\mathcal C$. This implies that for any $\bb x \in \partial C$, the ray starting from $\bb 0$ and passing through $\bb x$ has no other intersections with $\partial\mathcal C$: otherwise, the ray would need to exit and re-enter $\mathcal C$. In turn, this means that any $\bb x\in\partial\mathcal C$ has unique angular coordinates. Then the projection map $\bb\pi\colon\bb\varphi(\partial\mathcal C)\to S^2$ given by $\bb\pi(\bb x) = \bb x/\|\bb x\|$ is smooth and bijective, so we can define $\mathcal F \equiv (\bb\pi\circ\bb\varphi)^{-1}$.

Such an inflated cow can in fact be produced computationally by flowing points in the direction of increasing distance from the cow. To make this precise, we define the signed distance function $d_{\mathcal C}\colon\mathbb R^3 \to \mathbb R$ by
\begin{equation}
    d_{\mathcal C}(\bb x) = 
        \left(\min_{\bb y\in\partial\mathcal C}\|\bb x - \bb y \|\right)
        \times
        \begin{cases}
            -1 & \bb x\in\mathcal C \\
            1 & \bb x\notin\mathcal C.
        \end{cases}
\end{equation}
Now, observe that $\lim_{\|\bb x\|\to\infty}d_{\mathcal C}(\bb x)/\|\bb x\| = 1$, so at large distances, the level sets of $d_{\mathcal C}$ approach a dilated $S^2$, and thus they become trivially star-shaped. This behavior is demonstrated in \cref{fig:puffy-cow}. Now consider the gradient flow of $d_{\mathcal C}$, i.e., the function $\bb\varphi\colon{\mathbb R^3\times\mathbb R}\to\mathbb R^3$ satisfying $\partial_t\bb\varphi(\bb x, t) = \nabla d_{\mathcal C}(\bb\varphi(\bb x, t))$ with $\bb\varphi(\bb x, 0) = \bb x$. This function translates the point $\bb x$ along integral curves associated with the gradient of $d_{\mathcal C}$, orthogonal to the level sets. At large $t$, the function $\bb\varphi(\cdot, t)$ maps $\mathcal C$ to a star-shaped domain, and the inverse is given by $\bb\varphi(\cdot, -t)$. The resulting mapping to the unit sphere is illustrated in projection in \cref{fig:distance-gradient-flow}.

Using this map to define $\mathcal F$, we can then expand the components $(f_r, f_{\Delta\theta}, f_{\Delta\phi})$ in spherical harmonics. The first few multipoles are enumerated in \cref{tab:distance-gradient-flow-components}. The magnitude of the monopole term relative to higher-order terms gives us the first quantitative assessment of the reliability of the spherical cow approximation. The resulting geometry $\partial\mathcal C$ is shown at progressively higher multipole order in \cref{fig:gradient-flow-multipoles}.

\begin{figure}\centering
    \begin{tikzpicture}
        \foreach \i in {0,...,2} {
            \foreach \j in {0,...,2} {
                \node at (\j*3, -\i*2.6) {
                    \includegraphics[width=2.5cm]
                        {gradient_flow/grid_\i_\j.png}
                };
            }
        }
    \end{tikzpicture}
    \caption{Cow geometry $\partial\mathcal C$ reconstructed via the distance gradient flow method including progressively higher multipole contributions.\ \textbf{Top:} monopole, dipole, and quadrupole.\ \textbf{Middle:} octupole, hexadecapole, and dotriacontapole.\ \textbf{Bottom:} tetrahexacontapole, octacosahectapole, and full cow for comparison.}
    \label{fig:gradient-flow-multipoles}
\end{figure}

While the distance gradient flow method has an appealingly simple geometric interpretation, it is far from the simplest approach to implement. The flow tends to squeeze points along the ``sutures'' of the inflated cow (see \cref{fig:puffy-cow}), meaning that the integral curves must be computed with very high precision for many points. All that is really needed is for the cow to become \emph{rounder.} While the distance flow is an extremely simple way to accomplish that, there are more sophisticated techniques to more smoothly round out a cow, although they suffer from the clear deficiency that they do not produce a cow that looks as though it has been inflated with a bicycle pump. For example, one could use a smoothing operator to asymptotically map $\partial\mathcal C$ towards a surface with constant curvature, i.e., the sphere. Ultimately, however, a more established approach is provided by the method of harmonic maps.

\begin{figure*}
    \includegraphics[width=0.95\textwidth]{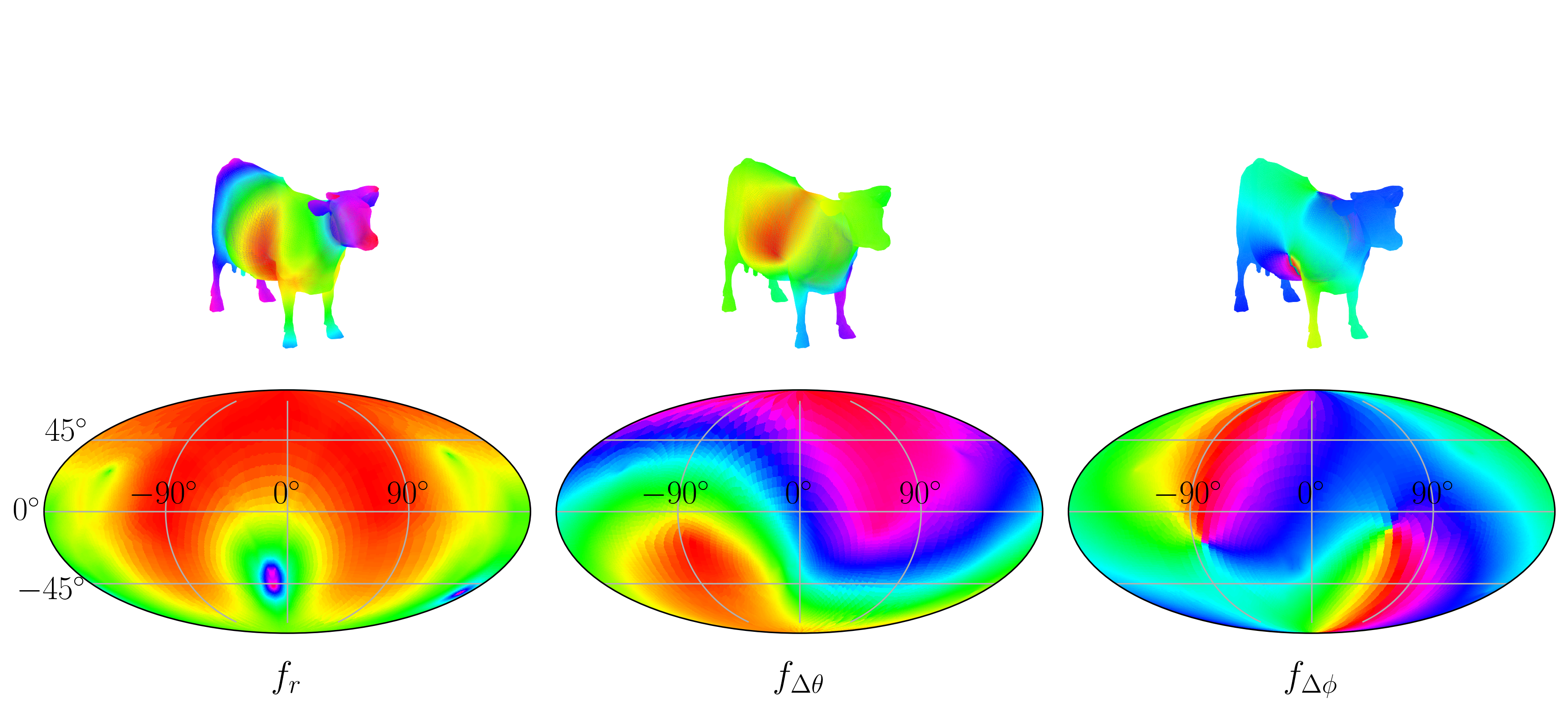}
    \caption{Homeomorphism $\mathcal F$ between the cow and the unit sphere obtained by the method of harmonic maps. In each column, colors correspond to the value of one component of $\bb f$. The unit sphere is represented via Mollweide projection in the bottom row.}
    \label{fig:harmonic-map-projections}
\end{figure*}

\subsection{Harmonic map method}
Another way to define a map from the cow to the sphere is to identify a set of harmonic coordinates on $\partial\mathcal C$. A major advantage of this strategy is that there exist powerful theorems that guarantee the features of the resulting map. In the previous section, we relied on ample handwaving to justify the smoothness and bijectivity of the map generated by the distance gradient flow, but in the harmonic map method, these properties can be easily proven.

A harmonic coordinate system on a manifold $M$ is a set of coordinate maps $x_i$ on $M$ such that each $x_i$ is harmonic, meaning that $\Delta x_i = 0$. Here $\Delta$ is the Laplace-Beltrami operator, a generalization of the Laplacian to Riemannian manifolds: just as the ordinary Laplacian $\nabla^2$ can be written as the divergence of the gradient of a scalar function on $\mathbb R^n$, the Laplace-Beltrami operator $\Delta$ is the Riemannian divergence of the Riemannian gradient of a scalar function on $M$. Explicitly, given a coordinate chart with a metric $g_{ij}$, the the Laplace-Beltrami operator takes the form~\cite{Flanders:1963gvr}
\begin{equation}
    \Delta f = \frac{1}{\sqrt{|g|}}\partial_i\left[
        \sqrt{|g|}g^{ij}\partial_j f
    \right]
    .
\end{equation}

It will turn out that if the map $\mathcal F^{-1}\colon \partial\mathcal C\to S^2$ is harmonic, then for sufficiently fine discretizations of $\partial\mathcal C$ and $S^2$, $\mathcal F$ is bijective without self-intersections of the mapped triangles. I refer the reader to \refcite{floater:2005} for the details. In particular, let us assume that $\partial\mathcal C$ is a genus-0 surface (again, omitting the alimentary canal in order to avoid stepping into truly messy calculations), and that the discretizations of the surfaces take the form of triangulations. For genus-0 surfaces, harmonic maps are equivalent to conformal maps~\cite{Gu:2002}, and there exists a substantial literature on computing discrete conformal maps between meshes, notably \refscite{Kharevych:2006,Springborn:2008}.

In practice, as discussed in the foregoing references, we can construct a harmonic map to the sphere by first deleting one vertex $v_0$ and the connecting faces, leaving a mesh with the topology of the disk. Efficient routines for computing harmonic maps between meshes with disk topology are widely available, and indeed, one such algorithm is implemented in \texttt{libigl.harmonic}. Once mapped harmonically to the circular disk, the mesh can be further mapped to the sphere via stereographic projection, which is also conformal. Finally, the deleted vertex $v_0$ is mapped to the pole of the stereographic projection, and the corresponding faces are restored. This fully specifies the homeomorphism $\mathcal F^{-1}\colon\partial\mathcal C\to\mathcal S^2$, and with it the inverse $\mathcal F$. This mapping is shown for each component separately in \cref{fig:harmonic-map-projections}.

\Cref{tab:harmonic-map-components} shows multipole coefficients for the map $\mathcal F$ constructed by this route. I selected the vertex for deletion in the middle of the cow's back, where the geometry is relatively smooth.\footnote{Experimental implementation of this procedure is strongly discouraged, as the deletion of vertices from live cows is highly unethical.} The basic structure of the multipole coefficients is very similar to those obtained by the distance gradient flow method (\cref{tab:distance-gradient-flow-components}). It is clear from \cref{fig:harmonic-map-projections} that most of the features in the map are at angular scales of $\mathcal O(\qty{10}{\degree})$ or larger. Each multipole captures features on angular scales of order $\pi/\ell$, so we should expect that the power spectrum will be negligible for $\ell \gtrsim 20$. This is indeed borne out by direct computation of these high multipoles. Of course, the present computation cannot predict very high multipoles $\ell \gg 100$ that correspond to angular scales smaller than the discretization scale of the benchmark cow---measuring these multipoles would require additional laboratory input.

\begin{table}\centering
    \caption{Multipole coefficients of the cow surface $\partial\mathcal C$ through the octupole in benchmark units, computed via the gradient flow method.}
    {\footnotesize
    \setlength{\tabcolsep}{3.5pt}
    \renewcommand{\arraystretch}{1.15}
    \setlength{\heavyrulewidth}{0.8pt}
    \setlength{\lightrulewidth}{0.5pt}
    \setlength{\cmidrulewidth}{0.5pt}
    \setlength{\aboverulesep}{0.4ex}
    \setlength{\belowrulesep}{0.4ex}
    \begin{tabular*}{\linewidth}{ccccc}
    \addlinespace[1ex]
    \toprule
    \addlinespace[-0.85ex]
    \toprule
    \addlinespace[1.5ex]
        $\ell$ & $m$ & $f_r$ & $f_{\Delta\theta}$ & $f_{\Delta\phi}$ \\
    \addlinespace[0.2ex]
    \midrule
    \addlinespace[1ex]
        0 & 0 &
             $1.0750$ &
             0 &
             0 \\
        1 & 0 &
             $0.0010$ &
             $0.7340$ &
             $0.0016$ \\
        1 & 1 &
             $0.0567 + 0.0145i$ &
             $-0.0008 + 0.0045i$ &
             $0.0236 - 0.0683i$ \\
        2 & 0 &
             $-0.2681$ &
             0.0051 &
             0.2863 \\
        2 & 1 &
             $-0.0007 - 0.0006i$ &
             $-0.0085 + 0.0474i$ &
             $0.0006 - 0.0045i$ \\
        2 & 2 &
             $0.1178 - 0.0438i$ &
             $0.0010 - 0.0006i$ &
             $-0.1602 - 0.0209i$ \\
        3 & 0 &
             $0.0008$ &
             $-0.1178$ &
             $-0.0017$ \\
        3 & 1 &
             $0.0182 - 0.0307i$ &
             $-0.0020 + 0.0028i$ &
             $0.0542 - 0.0645i$ \\
        3 & 2 &
             $-0.0003 + 0.0025i$ &
             $0.0489 + 0.0091i$ &
             $0.0001 + 0.0002i$ \\
        3 & 3 &
             $0.0392 + 0.0763i$ &
             $-0.0010 - 0.0011i$ &
             $-0.1240 + 0.0255i$ \\
    \bottomrule
    \addlinespace[0ex]
    \bottomrule
    \end{tabular*}
    }
    \label{tab:distance-gradient-flow-components}
\end{table}

\section{Bovine rigid body mechanics}
\label{sec:rigid-body}
Having now specified multiple schemes for defining the multipole coefficients of the cow's geometry, it is time to apply this expansion to an important real-world problem: cow tipping. The core problem of cow tipping is to determine the minimum amount of force that must be applied to a standing cow in order to cause it to fall onto its side horizontally. In our coordinates, the cow must rotate about the $x$ axis, and the force is applied in the $yz$ plane.

First, observe that this is a problem where the SCA completely fails. If the cow is spherical, and the coefficient of friction between the cow and the ground is large enough to prevent slipping, then the minimum force required to tip the cow is \emph{zero:} a very small force applied tangentially to the top of the cow will cause it to slowly roll until it has turned completely onto its side.

How much force is required to tip a cow with a general geometry? A cow will begin tipping when the normal force $\bb F_{\mathrm{N}}$ from the ground can no longer offset the torque from the tipping force. The magnitude and direction of $\bb F_{\mathrm{N}}$ are fixed by the weight of the cow less the $z$ component of the applied force, so the only parameter of the normal force that is free to vary is its effective point of application. There is some point $\bb p$ on the contact patch of the cow with the ground that maximizes the torque resulting from $\bb F_{\mathrm{N}}$ in the direction opposite to the torque $\bb\tau_{\mathrm{T}}$ from the tipping force. If the normal torque at this point is still less than $\tau_{\mathrm{T}}$, the cow will begin to tip, with $\bb p$ as the pivot point. This is why the spherical cow is easily tipped: there is only one point of contact with the ground, from which the normal force always points directly towards the barycenter, yielding no torque at all.

\begin{figure}\centering
    \includegraphics[width=0.45\textwidth]{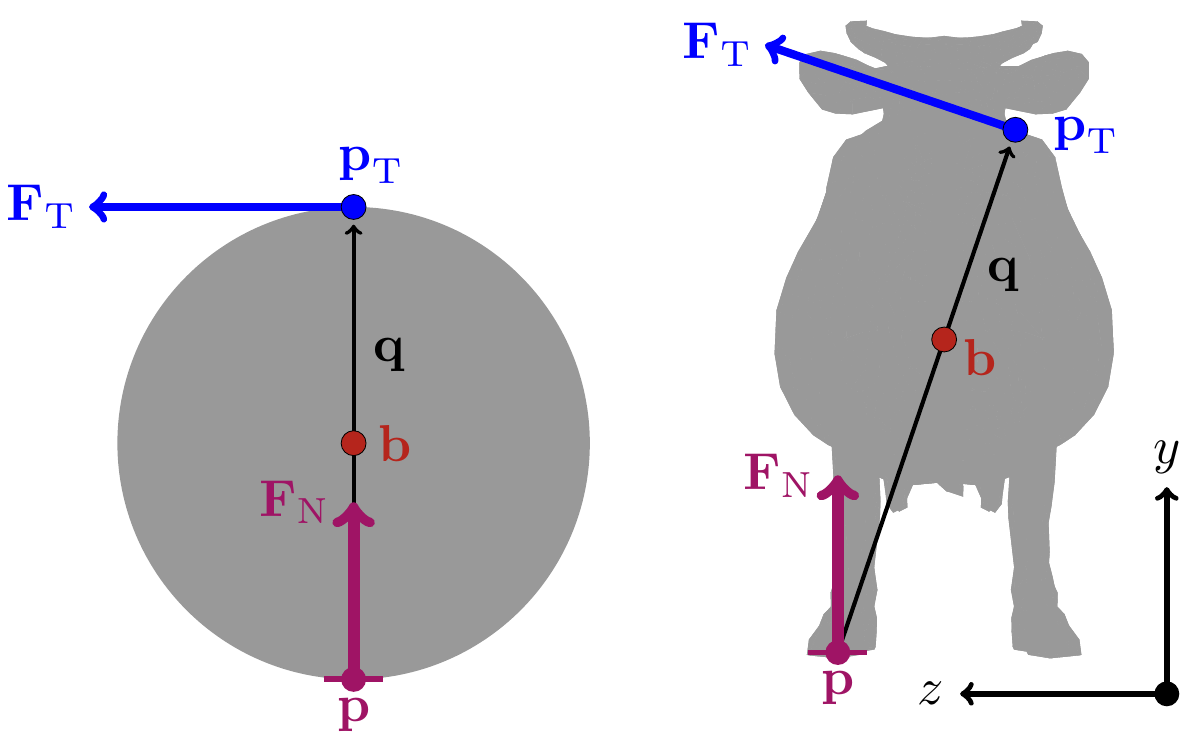}
    \caption{Geometry of the cow tipping problem. The tipping force is applied at $\bb p_{\mathrm{T}}$, and the compensating torque from the ground is applied at the contact patch, $\bb p$. The weight $\bb w$ is omitted from the diagram. \textbf{Left:} the tipping problem in the SCA. Any force is sufficient to tip the cow, since $\bb F_{\mathrm{N}}$ is parallel to $\bb q$. \textbf{Right:} the tipping problem for the realistic cow geometry. The normal force now generates a torque to resist tipping. Note that the cow is facing out of the page.}
    \label{fig:tipping}
\end{figure}

Let us assume that the tipping force $\bb F_{\mathrm{T}}$ is applied at the point $\bb p_{\mathrm{T}}$ at the intersection of $\partial\mathcal C$ with the ray $\bb q$ that originates at the pivot point on the ground and passes through the barycenter $\bb b$ of $\mathcal C$. In this case, the optimal direction for $\bb F_{\mathrm{T}}$ is the direction perpendicular to both $\bb q$ and $\bb{\hat x}$. Now, if $\bb F_{\mathrm{T}}\cdot\bb{\hat z} > 0$ (i.e., the cow is being tipped to the right), then $p_{\mathrm{T}}$ must be the point of contact with the ground with the largest value of $z$ (i.e., the furthest to the right). The arrangement of these vectors is illustrated in \cref{fig:tipping}. Given these data, it is now easy to compute the minimum force required to start the tipping process.

\begin{table}\centering
    \caption{Multipole coefficients of the cow surface $\partial\mathcal C$ through the quadrupole computed via the harmonic map method.}
    {\footnotesize
    \setlength{\tabcolsep}{3.5pt}
    \renewcommand{\arraystretch}{1.15}
    \setlength{\heavyrulewidth}{0.8pt}
    \setlength{\lightrulewidth}{0.5pt}
    \setlength{\cmidrulewidth}{0.5pt}
    \setlength{\aboverulesep}{0.4ex}
    \setlength{\belowrulesep}{0.4ex}
    \begin{tabular*}{\linewidth}{ccccc}
    \addlinespace[1ex]
    \toprule
    \addlinespace[-0.85ex]
    \toprule
    \addlinespace[1.5ex]
        $\ell$ & $m$ & $f_r$ & $f_{\Delta\theta}$ & $f_{\Delta\phi}$ \\
    \addlinespace[0.2ex]
    \midrule
    \addlinespace[1ex]
        0 & 0 &
             $0.5825$ &
             0 &
             0 \\
        1 & 0 &
             $-0.0996$ &
             $2.5793$ &
             $-0.4817$ \\
        1 & 1 &
             $0.0237 + 0.0000i$ &
             $-0.3161 + 1.6600i$ &
             $0.4750 + 0.4784i$ \\
        2 & 0 &
             $0.0106$ &
             $-0.0125$ &
             $-0.2309$ \\
        2 & 1 &
             $0.0193 + 0.0061i$ &
             $0.0425 - 0.2846i$ &
             $0.1581 + 1.5432i$ \\
        2 & 2 &
             $0.0591 + 0.0193i$ &
             $0.1197 - 0.1800i$ &
             $0.8938 - 0.0098i$ \\
        3 & 0 &
             $0.0005$ &
             $0.1995$ &
             $0.2132$ \\
        3 & 1 &
             $-0.0020 + 0.0003i$  &
             $0.0226 - 0.0683i $ &
             $0.1065 - 0.3695i $ \\
        3 & 2 &
             $-0.0358 - 0.0127i$ &
             $-0.0657 + 0.0821i$ &
             $-0.2082 - 0.0496i$ \\
        3 & 3 &
             $-0.0083 - 0.0094i$ &
             $-0.0272 + 0.0286i$ &
             $-0.1859 + 0.1900i$ \\
    \bottomrule
    \addlinespace[0ex]
    \bottomrule
    \end{tabular*}
    }
    \label{tab:harmonic-map-components}
\end{table}

If the force is sufficient to start tipping, and the contact patch with the ground does not change, then the same force is sufficient to complete the tip: as the tip begins, the barycenter moves in the positive $z$ direction while the pivot stays fixed, reducing the torque applied by the normal force and favoring further tipping. Thus, most prior analyses have simply computed the amount of force required to start the tipping process. This is found by equating the tipping torque $\tau_{\mathrm{T}} = F_{\mathrm{T}}\|\bb p_{\mathrm{T}} - \bb b\|$ to the torque at the pivot point, $F_{\mathrm{N}}\| (\bb p - \bb b)\times\bb{\hat y} \|$. The magnitude of the normal force is $F_{\mathrm{N}} = w - \bb F_{\mathrm{T}}\cdot\bb{\hat y}$, where $w$ is the weight of the cow.

It is certainly possible to implement exactly this procedure for the multipole expansion of the cow geometry at any given order. The $y$ coordinate is analytically minimized to determine a discrete set of candidate pivot points, and of these, the pivot $\bb p$ is the point with maximal $z$ coordinate. Once $\bb p$ is determined, one can solve e.g. for $\theta$ in $\theta + f_{\Delta\theta}(\theta) = p_\theta + \pi$ to find the coordinates on $S^2$ corresponding under $\mathcal F$ to the point $\bb p_{\mathrm T}$ at which the force is applied. The minimum required force $F_{\mathrm{T}}$ is then found by simple vector algebra, as above.

But the multipole expansion allows us to go further. Since it is simple to compute the inertia tensor of the cow at any order in the expansion, we can separately consider the case of a momentary impulse applied at $\bb p_{\mathrm{T}}$. It then becomes a simple exercise to determine whether the torque applied by the normal force will overcome that angular impulse before the center of mass moves to the other side of the pivot point, ensuring the completion of the tip. This is extremely important in the cow tipping problem, because the maximum \emph{sustainable} force that can be produced by a human is much less than the maximum \emph{momentary} force that can be applied. A typical human can sustain a pushing force of about \qty{500}{\newton}, which leads to the conclusion that cow tipping is well out of reach, by a factor of a few. However, elite boxers are known to punch with forces of order \qty{5000}{\newton}, larger by an order of magnitude. It is certainly conceivable that the impulse delivered by such a strong momentary burst could lead to a complete tip of the cow. While time-dependent data on human strength is available~\cite{nasa}, I defer a detailed analysis to future work.

\section{Discussion and conclusions}
\label{sec:discussion}
In the foregoing sections, I have introduced multiple schemes for interpreting the spherical cow as the first term of a \emph{multipole cow.} This allows for the first quantitative tests of the SCA, and I find that there are several cases in which the SCA fails badly. This work also significantly extends previous literature on bovine rigid body mechanics, particularly the cow tipping problem, where the SCA dramatically underestimates the force required. This improved theoretical understanding of cow tipping is essential, since experimental work on this subject faces significant ethical barriers.

The main virtue of the multipole expansion is that it is systematically improvable, with interpretable behavior at each successive multipole order. It is not necessarily the most economical or fastest-converging scheme for approximating the cow. Various claims have been made at times in the context of elephantine approximations, notably that four parameters are sufficient to fit an elephant, and that a fifth can be used to incorporate mobility of the tail~\cite{Dyson:2004}. More recent explorations have shown that it is indeed possible to find such fits~\cite{Jin:2024}. However, were these methods to be applied to the cow, they would still not provide an interpretable, systematically improvable approximation scheme.

One weakness of the numerical results presented here is that they correspond only to one benchmark cow. Physical cows exhibit individual variation that presents a fundamental limit to precision bovine modeling. In this work, I have generally limited results to the first few multipoles, which are likely to be fairly robust, but I have made no effort to quantify the distribution of each of the multipole coefficients across the bovine population. It is also worth noting that results may vary for different species of cow, and here I have considered only \textit{Bos taurus}.

Additionally, while the multipolar expansion performed in this work can reproduce the \textit{shape} of the cow, there are several other considerations for bovine mechanics that require an independent set of approximation schemes. In particular, I have assumed that the cow is entirely homogeneous, with uniform density. This is well known not to be the case. Even neglecting bone structure, a significant fraction of the interior volume of the cow is occupied by the stomach, whose largest component can displace up to \qty{200}{\liter}~\cite{Milo:2009}. Inhomogeneity has significant implications for the cow tipping problem, and any reliable treatment must include these effects.

The treatment in this work also provides no diagnostic of compressibility and elasticity, which are important for a complete treatment of gravitational radiation at higher angular frequencies: for $\omega \gg \qty{1}{\hertz}$, centrifugal forces distort the shape of the cow, modifying the emitted power. While this distortion contributes most significantly to the dipole, which is unimportant for gravitational radiation, all higher multipoles will also be affected. I stress, however, that for angular frequencies at which one might realistically find a cow spinning in space, these are extremely small effects. Moreover, the framework developed here can be readily extended to incorporate oblateness, exactly as is done for the shape of Earth: the reference EGM2008 model for Earth's shape is itself defined in terms of spherical harmonics~\cite{2012JGRB..117.4406P}.

Ultimately, the SCA is quite accurate in many physical problems with rough spherical symmetry. However, in more general circumstances, the spherical cow approximation risks producing nothing but bull. This is not to say that the SCA should be put out to pasture. Rather, I would suggest only that when choosing tools for a given problem, it is sometimes best not to follow the herd.

\begin{acknowledgments}
I thank Innes Bigaran, Sarah Geller, Natasha Keces, Lee Rosenthal, and Aaron Vincent for valuable discussions, and I acknowledge alarming current events for driving me to the distraction of investigating this problem. No cows were harmed in the course of this work, and no Dehn surgery was performed without veterinary supervision. I declare no conflicts of interest (apart from vegetarianism). Although my work has been generally supported by the MIT Pappalardo Fellowship and Center for Theoretical Physics---a Leinweber Institute, this work was performed on my own time, and definitely incurred no cost to the US government. The reader who still objects to this use of my off-hours is invited to check the date---and to not have a cow, man.
\end{acknowledgments}

The author has no conflicts to disclose.

\bibliography{references}  

\begin{thebibliography}{19}%
\makeatletter
\providecommand \@ifxundefined [1]{%
 \@ifx{#1\undefined}
}%
\providecommand \@ifnum [1]{%
 \ifnum #1\expandafter \@firstoftwo
 \else \expandafter \@secondoftwo
 \fi
}%
\providecommand \@ifx [1]{%
 \ifx #1\expandafter \@firstoftwo
 \else \expandafter \@secondoftwo
 \fi
}%
\providecommand \natexlab [1]{#1}%
\providecommand \enquote  [1]{``#1''}%
\providecommand \bibnamefont  [1]{#1}%
\providecommand \bibfnamefont [1]{#1}%
\providecommand \citenamefont [1]{#1}%
\providecommand \href@noop [0]{\@secondoftwo}%
\providecommand \href [0]{\begingroup \@sanitize@url \@href}%
\providecommand \@href[1]{\@@startlink{#1}\@@href}%
\providecommand \@@href[1]{\endgroup#1\@@endlink}%
\providecommand \@sanitize@url [0]{\catcode `\\12\catcode `\$12\catcode
  `\&12\catcode `\#12\catcode `\^12\catcode `\_12\catcode `\%12\relax}%
\providecommand \@@startlink[1]{}%
\providecommand \@@endlink[0]{}%
\providecommand \url  [0]{\begingroup\@sanitize@url \@url }%
\providecommand \@url [1]{\endgroup\@href {#1}{\urlprefix }}%
\providecommand \urlprefix  [0]{URL }%
\providecommand \Eprint [0]{\href }%
\providecommand \doibase [0]{https://doi.org/}%
\providecommand \selectlanguage [0]{\@gobble}%
\providecommand \bibinfo  [0]{\@secondoftwo}%
\providecommand \bibfield  [0]{\@secondoftwo}%
\providecommand \translation [1]{[#1]}%
\providecommand \BibitemOpen [0]{}%
\providecommand \bibitemStop [0]{}%
\providecommand \bibitemNoStop [0]{.\EOS\space}%
\providecommand \EOS [0]{\spacefactor3000\relax}%
\providecommand \BibitemShut  [1]{\csname bibitem#1\endcsname}%
\let\auto@bib@innerbib\@empty
\bibitem [{sch()}]{scholar}%
  \BibitemOpen
  \href@noop {} {\enquote {\bibinfo {title} {{``spherical cow'' - Google
  Scholar}},}\ }\bibinfo {note}
  {\\\href{https://scholar.google.com/scholar?q=\%22spherical+cow\%22}{scholar.google.com/scholar?q=\%22spherical+cow\%22}\\(accessed
  March 2025)}\BibitemShut {NoStop}%
\bibitem [{\citenamefont {Jacobson}, \citenamefont {Panozzo}\ \emph
  {et~al.}(2018{\natexlab{a}})\citenamefont {Jacobson}, \citenamefont {Panozzo}
  \emph {et~al.}}]{libigl}%
  \BibitemOpen
  \bibfield  {author} {\bibinfo {author} {\bibfnamefont {A.}~\bibnamefont
  {Jacobson}}, \bibinfo {author} {\bibfnamefont {D.}~\bibnamefont {Panozzo}},
  \emph {et~al.},\ }\href@noop {} {\enquote {\bibinfo {title}
  {{\texttt{libigl}}: A simple {C++} geometry processing library},}\ }
  (\bibinfo {year} {2018}{\natexlab{a}}),\ \bibinfo {note}
  {\href{https://libigl.github.io/}{libigl.github.io}}\BibitemShut {NoStop}%
\bibitem [{\citenamefont {Jacobson}, \citenamefont {Panozzo}\ \emph
  {et~al.}(2018{\natexlab{b}})\citenamefont {Jacobson}, \citenamefont {Panozzo}
  \emph {et~al.}}]{libigl-cow}%
  \BibitemOpen
  \bibfield  {author} {\bibinfo {author} {\bibfnamefont {A.}~\bibnamefont
  {Jacobson}}, \bibinfo {author} {\bibfnamefont {D.}~\bibnamefont {Panozzo}},
  \emph {et~al.},\ }\href@noop {} {\enquote {\bibinfo {title}
  {\texttt{cow.off}},}\ } (\bibinfo {year} {2018}{\natexlab{b}}),\ \bibinfo
  {note}
  {\\\href{https://github.com/libigl/libigl-tutorial-data/blob/master/cow.off}{github.com/libigl/libigl-tutorial-data/blob/master/cow.off}
  (accessed March 2025)}\BibitemShut {NoStop}%
\bibitem [{\citenamefont {Favre}\ and\ \citenamefont
  {Powell}(2014)}]{IUPAC:2014}%
  \BibitemOpen
  \bibfield  {author} {\bibinfo {author} {\bibfnamefont {H.~A.}\ \bibnamefont
  {Favre}}\ and\ \bibinfo {author} {\bibfnamefont {W.~H.}\ \bibnamefont
  {Powell}},\ }\href {https://doi.org/10.1039/9781849733069} {\emph {\bibinfo
  {title} {{Nomenclature of Organic Chemistry: IUPAC Recommendations and
  Preferred Names 2013}}}}\ (\bibinfo  {publisher} {RSC Publishing},\ \bibinfo
  {year} {2014})\BibitemShut {NoStop}%
\bibitem [{\citenamefont {Lott}(1981)}]{lott:1981}%
  \BibitemOpen
  \bibfield  {author} {\bibinfo {author} {\bibfnamefont {D.~F.}\ \bibnamefont
  {Lott}},\ }\bibfield  {title} {\enquote {\bibinfo {title} {{Sexual behavior
  and intersexual strategies in American bison}},}\ }\href
  {https://doi.org/10.1111/j.1439-0310.1981.tb01289.x} {\bibfield  {journal}
  {\bibinfo  {journal} {Zeitschrift f{\"u}r Tierpsychologie}\ }\textbf
  {\bibinfo {volume} {56}},\ \bibinfo {pages} {97--114} (\bibinfo {year}
  {1981})}\BibitemShut {NoStop}%
\bibitem [{\citenamefont {Douglass}(1999)}]{douglass:1999}%
  \BibitemOpen
  \bibfield  {author} {\bibinfo {author} {\bibfnamefont {C.~B.}\ \bibnamefont
  {Douglass}},\ }\href {https://doi.org/10.2307/j.ctv2jhjw0x} {\emph {\bibinfo
  {title} {Bulls, bullfighting, and Spanish identities}}}\ (\bibinfo
  {publisher} {University of Arizona Press},\ \bibinfo {year}
  {1999})\BibitemShut {NoStop}%
\bibitem [{\citenamefont {Murphy}\ \emph {et~al.}(2010)\citenamefont {Murphy},
  \citenamefont {McGuire}, \citenamefont {O’Malley},\ and\ \citenamefont
  {Harrington}}]{murphy:2010}%
  \BibitemOpen
  \bibfield  {author} {\bibinfo {author} {\bibfnamefont {C.~G.}\ \bibnamefont
  {Murphy}}, \bibinfo {author} {\bibfnamefont {C.~M.}\ \bibnamefont {McGuire}},
  \bibinfo {author} {\bibfnamefont {N.}~\bibnamefont {O’Malley}},\ and\
  \bibinfo {author} {\bibfnamefont {P.}~\bibnamefont {Harrington}},\ }\bibfield
   {title} {\enquote {\bibinfo {title} {{Cow-related trauma: a 10-year review
  of injuries admitted to a single institution}},}\ }\href
  {https://doi.org/10.1016/j.injury.2009.08.006} {\bibfield  {journal}
  {\bibinfo  {journal} {Injury}\ }\textbf {\bibinfo {volume} {41}},\ \bibinfo
  {pages} {548--550} (\bibinfo {year} {2010})}\BibitemShut {NoStop}%
\bibitem [{\citenamefont {Jackson}(1998)}]{Jackson:1998nia}%
  \BibitemOpen
  \bibfield  {author} {\bibinfo {author} {\bibfnamefont {J.~D.}\ \bibnamefont
  {Jackson}},\ }\href@noop {} {\emph {\bibinfo {title} {{Classical
  Electrodynamics}}}}\ (\bibinfo  {publisher} {Wiley},\ \bibinfo {year}
  {1998})\BibitemShut {NoStop}%
\bibitem [{\citenamefont {Maggiore}(2007)}]{Maggiore:2007ulw}%
  \BibitemOpen
  \bibfield  {author} {\bibinfo {author} {\bibfnamefont {M.}~\bibnamefont
  {Maggiore}},\ }\href
  {https://doi.org/10.1093/acprof:oso/9780198570745.001.0001} {\emph {\bibinfo
  {title} {{Gravitational Waves. Vol. 1: Theory and Experiments}}}}\ (\bibinfo
  {publisher} {Oxford University Press},\ \bibinfo {year} {2007})\BibitemShut
  {NoStop}%
\bibitem [{\citenamefont {Flanders}(1963)}]{Flanders:1963gvr}%
  \BibitemOpen
  \bibfield  {author} {\bibinfo {author} {\bibfnamefont {H.}~\bibnamefont
  {Flanders}},\ }\href@noop {} {\emph {\bibinfo {title} {{Differential Forms
  with Applications to the Physical Sciences}}}},\ \bibinfo {series}
  {Mathematics in Science and Engineering}, Vol.~\bibinfo {volume} {11}\
  (\bibinfo  {publisher} {Academic Press},\ \bibinfo {year} {1963})\BibitemShut
  {NoStop}%
\bibitem [{\citenamefont {Floater}\ and\ \citenamefont
  {Hormann}(2005)}]{floater:2005}%
  \BibitemOpen
  \bibfield  {author} {\bibinfo {author} {\bibfnamefont {M.~S.}\ \bibnamefont
  {Floater}}\ and\ \bibinfo {author} {\bibfnamefont {K.}~\bibnamefont
  {Hormann}},\ }\bibfield  {title} {\enquote {\bibinfo {title} {Surface
  parameterization: a tutorial and survey},}\ }in\ \href
  {https://doi.org/10.1007/3-540-26808-1_9} {\emph {\bibinfo {booktitle}
  {Advances in Multiresolution for Geometric Modelling}}},\ \bibinfo {editor}
  {edited by\ \bibinfo {editor} {\bibfnamefont {N.~A.}\ \bibnamefont
  {Dodgson}}, \bibinfo {editor} {\bibfnamefont {M.~S.}\ \bibnamefont
  {Floater}},\ and\ \bibinfo {editor} {\bibfnamefont {M.~A.}\ \bibnamefont
  {Sabin}}}\ (\bibinfo  {publisher} {Springer Berlin Heidelberg},\ \bibinfo
  {address} {Berlin, Heidelberg},\ \bibinfo {year} {2005})\ pp.\ \bibinfo
  {pages} {157--186}\BibitemShut {NoStop}%
\bibitem [{\citenamefont {Gu}\ and\ \citenamefont {Yau}(2002)}]{Gu:2002}%
  \BibitemOpen
  \bibfield  {author} {\bibinfo {author} {\bibfnamefont {X.}~\bibnamefont
  {Gu}}\ and\ \bibinfo {author} {\bibfnamefont {S.}~\bibnamefont {Yau}},\
  }\bibfield  {title} {\enquote {\bibinfo {title} {Computing conformal
  structure of surfaces},}\ }\href {http://arxiv.org/abs/cs/0212043} {\bibfield
   {journal} {\bibinfo  {journal} {CoRR}\ }\textbf {\bibinfo {volume}
  {cs.GR/0212043}} (\bibinfo {year} {2002})}\BibitemShut {NoStop}%
\bibitem [{\citenamefont {Kharevych}, \citenamefont {Springborn},\ and\
  \citenamefont {Schr\"{o}der}(2006)}]{Kharevych:2006}%
  \BibitemOpen
  \bibfield  {author} {\bibinfo {author} {\bibfnamefont {L.}~\bibnamefont
  {Kharevych}}, \bibinfo {author} {\bibfnamefont {B.}~\bibnamefont
  {Springborn}},\ and\ \bibinfo {author} {\bibfnamefont {P.}~\bibnamefont
  {Schr\"{o}der}},\ }\bibfield  {title} {\enquote {\bibinfo {title} {Discrete
  conformal mappings via circle patterns},}\ }\href
  {https://doi.org/10.1145/1138450.1138461} {\bibfield  {journal} {\bibinfo
  {journal} {ACM Trans. Graph.}\ }\textbf {\bibinfo {volume} {25}},\ \bibinfo
  {pages} {412–438} (\bibinfo {year} {2006})}\BibitemShut {NoStop}%
\bibitem [{\citenamefont {Springborn}, \citenamefont {Schr\"{o}der},\ and\
  \citenamefont {Pinkall}(2008)}]{Springborn:2008}%
  \BibitemOpen
  \bibfield  {author} {\bibinfo {author} {\bibfnamefont {B.}~\bibnamefont
  {Springborn}}, \bibinfo {author} {\bibfnamefont {P.}~\bibnamefont
  {Schr\"{o}der}},\ and\ \bibinfo {author} {\bibfnamefont {U.}~\bibnamefont
  {Pinkall}},\ }\bibfield  {title} {\enquote {\bibinfo {title} {Conformal
  equivalence of triangle meshes},}\ }in\ \href
  {https://doi.org/10.1145/1399504.1360676} {\emph {\bibinfo {booktitle} {ACM
  SIGGRAPH 2008 Papers}}},\ \bibinfo {series and number} {SIGGRAPH '08}\
  (\bibinfo  {publisher} {Association for Computing Machinery},\ \bibinfo
  {address} {New York, NY, USA},\ \bibinfo {year} {2008})\BibitemShut {NoStop}%
\bibitem [{nas()}]{nasa}%
  \BibitemOpen
  \href@noop {} {\enquote {\bibinfo {title} {{HUMAN PERFORMANCE
  CAPABILITIES}},}\ }\bibinfo {note}
  {\\\href{https://web.archive.org/web/20250304202426/msis.jsc.nasa.gov/sections/section04.htm\#\_4.9\_STRENGTH}{web.archive.org/web/20250304202426/msis.jsc.nasa.gov\\\strut~~~~/sections/section04.htm\#\_4.9\_STRENGTH}\\(accessed
  March 2025)}\BibitemShut {NoStop}%
\bibitem [{\citenamefont {Dyson}(2004)}]{Dyson:2004}%
  \BibitemOpen
  \bibfield  {author} {\bibinfo {author} {\bibfnamefont {F.}~\bibnamefont
  {Dyson}},\ }\bibfield  {title} {\enquote {\bibinfo {title} {{A meeting with
  Enrico Fermi}},}\ }\href {https://doi.org/10.1038/427297a} {\bibfield
  {journal} {\bibinfo  {journal} {Nature}\ }\textbf {\bibinfo {volume} {427}},\
  \bibinfo {pages} {297--297} (\bibinfo {year} {2004})}\BibitemShut {NoStop}%
\bibitem [{\citenamefont {Jin}\ and\ \citenamefont {Yuan}(2024)}]{Jin:2024}%
  \BibitemOpen
  \bibfield  {author} {\bibinfo {author} {\bibfnamefont {D.}~\bibnamefont
  {Jin}}\ and\ \bibinfo {author} {\bibfnamefont {J.}~\bibnamefont {Yuan}},\
  }\bibfield  {title} {\enquote {\bibinfo {title} {Fitting an elephant with
  four non-zero parameters},}\ }\href@noop {} {\bibfield  {journal} {\bibinfo
  {journal} {arXiv}\ } (\bibinfo {year} {2024})},\ \Eprint
  {https://arxiv.org/abs/2407.07909} {2407.07909} \BibitemShut {NoStop}%
\bibitem [{\citenamefont {Milo}\ \emph {et~al.}(2009)\citenamefont {Milo},
  \citenamefont {Jorgensen}, \citenamefont {Moran}, \citenamefont {Weber},\
  and\ \citenamefont {Springer}}]{Milo:2009}%
  \BibitemOpen
  \bibfield  {author} {\bibinfo {author} {\bibfnamefont {R.}~\bibnamefont
  {Milo}}, \bibinfo {author} {\bibfnamefont {P.}~\bibnamefont {Jorgensen}},
  \bibinfo {author} {\bibfnamefont {U.}~\bibnamefont {Moran}}, \bibinfo
  {author} {\bibfnamefont {G.}~\bibnamefont {Weber}},\ and\ \bibinfo {author}
  {\bibfnamefont {M.}~\bibnamefont {Springer}},\ }\bibfield  {title} {\enquote
  {\bibinfo {title} {Bionumbers---the database of key numbers in molecular and
  cell biology},}\ }\href {https://doi.org/10.1093/nar/gkp889} {\bibfield
  {journal} {\bibinfo  {journal} {Nucleic Acids Research}\ }\textbf {\bibinfo
  {volume} {38}},\ \bibinfo {pages} {D750--D753} (\bibinfo {year} {2009})},\
  \bibinfo {note} {{BNID 117319}}\BibitemShut {NoStop}%
\bibitem [{\citenamefont {{Pavlis}}\ \emph {et~al.}(2012)\citenamefont
  {{Pavlis}}, \citenamefont {{Holmes}}, \citenamefont {{Kenyon}},\ and\
  \citenamefont {{Factor}}}]{2012JGRB..117.4406P}%
  \BibitemOpen
  \bibfield  {author} {\bibinfo {author} {\bibfnamefont {N.~K.}\ \bibnamefont
  {{Pavlis}}}, \bibinfo {author} {\bibfnamefont {S.~A.}\ \bibnamefont
  {{Holmes}}}, \bibinfo {author} {\bibfnamefont {S.~C.}\ \bibnamefont
  {{Kenyon}}},\ and\ \bibinfo {author} {\bibfnamefont {J.~K.}\ \bibnamefont
  {{Factor}}},\ }\bibfield  {title} {\enquote {\bibinfo {title} {{The
  development and evaluation of the Earth Gravitational Model 2008
  (EGM2008)}},}\ }\href {https://doi.org/10.1029/2011JB008916} {\bibfield
  {journal} {\bibinfo  {journal} {Journal of Geophysical Research (Solid
  Earth)}\ }\textbf {\bibinfo {volume} {117}},\ \bibinfo {eid} {B04406}
  (\bibinfo {year} {2012})}\BibitemShut {NoStop}%
\end{thebibliography}%

\end{document}